# Real-time 3D Nanoscale Coherent Imaging via Physics-aware Deep Learning

Henry Chan[1,6,*], Youssef S.G. Nashed[2], Saugat Kandel[3], Stephan Hruszkewycz[4], Subramanian Sankaranarayanan[1,6], Ross J. Harder[5] and Mathew J. Cherukara,[1,*]

[1]Center for Nanoscale Materials, Argonne National Laboratory, Argonne, IL, 60439
[2]Stats Perform, Chicago, IL, 60601
[3]Applied Physics, Northwestern University, Evanston, IL 60208
[4]Materials Science Division, Argonne National Laboratory, Argonne, IL, 60439
[5]Advanced Photon Source, Argonne National Laboratory, Argonne, IL, 60439
[6]Department of Mechanical and Industrial Engineering, University of Illinois, Chicago, IL 60607

## Abstract

**Phase retrieval, the problem of recovering lost phase information from measured intensity alone, is an inverse problem that is widely faced in various imaging modalities ranging from astronomy to nanoscale imaging. The current process of phase recovery is iterative in nature. As a result, the image formation is time-consuming and computationally expensive, precluding real-time imaging. Here, we use 3D nanoscale X-ray imaging as a representative example to develop a deep learning model to address this phase retrieval problem. We introduce 3D-CDI-NN, a deep convolutional neural network and differential programming framework trained to predict 3D structure and strain solely from input 3D X-ray coherent scattering data. Our networks are designed to be 'physics-aware' in multiple aspects; in that the physics of x-ray scattering process is explicitly enforced in the training of the network, and the training data are drawn from atomistic simulations that are representative of the physics of the material. We further refine the neural network prediction through a physics-based optimization procedure to enable maximum accuracy at lowest computational cost. 3D-CDI-NN can invert a 3D coherent diffraction pattern to real-space structure and strain hundreds of times faster than traditional iterative phase retrieval methods, with negligible loss in accuracy. Our integrated machine learning and differential programming solution to the phase retrieval problem is broadly applicable across inverse problems in other application areas.**

[*] mcherukara@anl.gov, hchan@anl.gov

# Introduction

Phase retrieval, which is the problem of recovering lost phases from measured intensities alone is the underlying basis for a variety of imaging modalities in astronomy,[1] Lorentz transmission electron microscopy (Lorentz-TEM),[2] super-resolution optical imaging,[3] and of particular relevance for this article, electron and X-ray coherent diffraction imaging (CDI) techniques including Ptychographic methods.[4,5] In CDI, for instance, the object of interest is illuminated with a coherent beam and the resulting scattered intensities are measured in the far-field. In the purest form, these measured intensities correspond to the modulus of the complex Fourier transform of the measured sample. While scattered intensities can be measured, the phase information contained in the scattered wavefield is lost. Consequently, the image cannot be retrieved with a simple inverse Fourier Transform.

Coherent imaging techniques are acutely sensitive to material properties that influence the phase of the scattered wave. When measured at a Bragg peak, local distortions of the lattice within the sample will directly impact the relative phases within the scattered wavefield. The coherent diffraction interference pattern will then encode the lattice distortion within the sample.[6] Recovering the object structure (and hence also phase) from the scattered intensities provides a 3D image of both the object's structure as well as the distortion of the lattice (represented as relative phase within the complex image) with sensitivity on the order of a few picometers.[7] This ability to obtain nanoscale structure and picometer sensitivity to distortions caused by strain has been widely used by the materials and chemistry communities to study a variety of dynamic processes resolved in time using X-ray Bragg CDI. Some examples include grain growth and annealing,[8] defect migration in battery electrodes,[9] ultra-fast phonon dynamics,[10–12] *in-situ* catalysis[13,14] and mechanical deformation.[15,16] While X-ray CDI has grown to be an extremely powerful means of characterizing the *in-situ* and *operando* response of materials, the process of phase retrieval is computationally expensive. Iterative phase retrieval methods typically require thousands of iterations and multiple runs from random initialization to converge to a solution of high accuracy, taking several minutes even on modern graphical processing units (GPU). Furthermore, convergence of the algorithms is sensitive to optimization hyper-parameters such as the choice of algorithms, algorithmic parameters, data thresholds, and data initialization.[17,18] These challenges preclude real-time phase retrieval and feedback, which is highly desirable across a broad range of imaging modalities.

Neural network solutions have been proposed to quickly solve various inverse problems including in magnetic resonance imaging (MRI),[19] inverse design of opto-electronic devices,[20] and phase retrieval problems.[21,22] While these results show the promise of deep learning in providing rapid solutions, more general concerns about the susceptibility of neural networks to sudden failures remain. These include their inability to extrapolate and generalize to inputs outside of the training distribution and their susceptibility to subtle biases in the training data. For instance, it was shown that a deep neural network that was approved for use as a medical device in Europe to detect skin melanomas was often making its predictions based on the presence of surgical markers in the dermoscopic images and not from any skin features.[23] What is needed then is a means of correcting predictions from neural networks in the event of errors, regardless of their magnitude. Additionally, while generative models have been widely applied to generate 2D images, generation of 3D structures is a nascent field.[24] The data requirements to model 3D structures are larger than for 2D, and the addition of an extra dimension means that there are more symmetries that need to be learned.

Here, we introduce a framework that uses a 3D convolutional encoder-decoder network (3D-CDI-NN) in conjunction with a physics-based optimization procedure to solve the inverse problem in 3D, using coherent imaging as a representative example. We use the reverse-mode automatic differentiation (AD) method both to make the 3D-CDI-NN model *physics-aware* during the training phase, and to refine the predicted image during the testing phase. We demonstrate that such an integrated approach of using a physics-based refinement stage on the 3D-CDI-NN prediction maximizes the speed and the accuracy of the inversion procedure. Our approach is applicable to several inverse problems and only requires knowledge of the forward model, where both the training data set and the refinement through optimization are derived.

## Results

**Approach.** Figure 1 illustrates our approach for inverting 3D coherent imaging data to real-space structure and strain field. The workflow consists of two stages: first, there is a computationally intensive offline training stage that involves training the 3D-CDI-NN model on data generated from large-scale atomistic simulations. Second, the trained 3D-CDI-NN is used in a fast online prediction stage that enables real-time predictions of 3D structure and strain. These predictions can then be refined using a gradient-based optimization procedure such as automatic differentiation.[25]

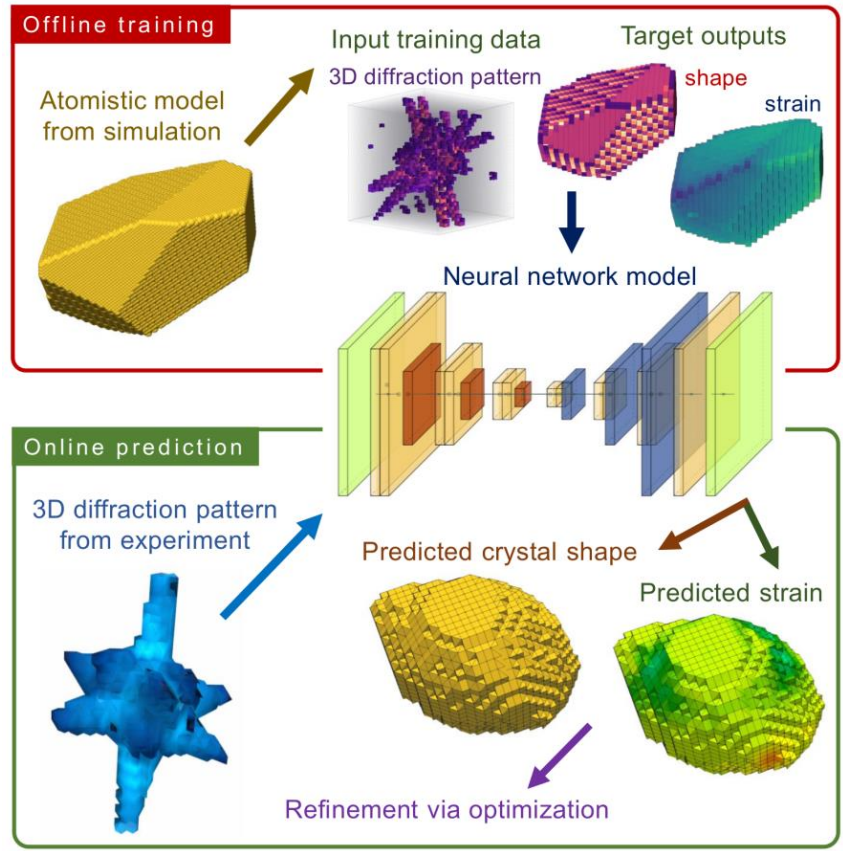

*Fig. 1: Schematic of physics-aware framework for phase retrieval in 3D coherent diffraction imaging.* *The main component of the framework is a neural network model (3D-CDI-NN) that is trained offline using 3D data (simulated diffraction pattern, crystal shape and local strain) derived from atomistic simulations that capture physics of the material. Once trained, the 3D-CDI-NN model can perform real-time prediction of crystal shape and local strain from experimentally measured diffraction pattern. The prediction can then be refined using a gradient-based optimization procedure.*

**Physics informed training set.** Effective training of a neural network hinges on the availability of training data that is sufficiently diverse and representative. To obtain training data that is representative of experimental data, we derive them from a physics informed data preparation pipeline using atomistic structures (Fig. 2). Each example in the training set is created as follows: First, a polyhedral shape is generated by clipping a cube shaped (FCC lattice) crystal along randomly selected high-symmetry orientations (see Methods). A random combination of compression, tension and shear stresses is applied on the atomistic object to create a strain field in the material. The structure is then energetically relaxed using LAMMPS,[26] a parallel molecular dynamics (MD) simulation package. This energy minimized atomic configuration is then used to calculate atom densities and displacements, which are spatially voxelized into a 32×32×32 grid (length of each voxel is ~ 2 lattice units) and used to compute the 3D coherent diffraction patterns about the (111) Bragg peak (see Methods).

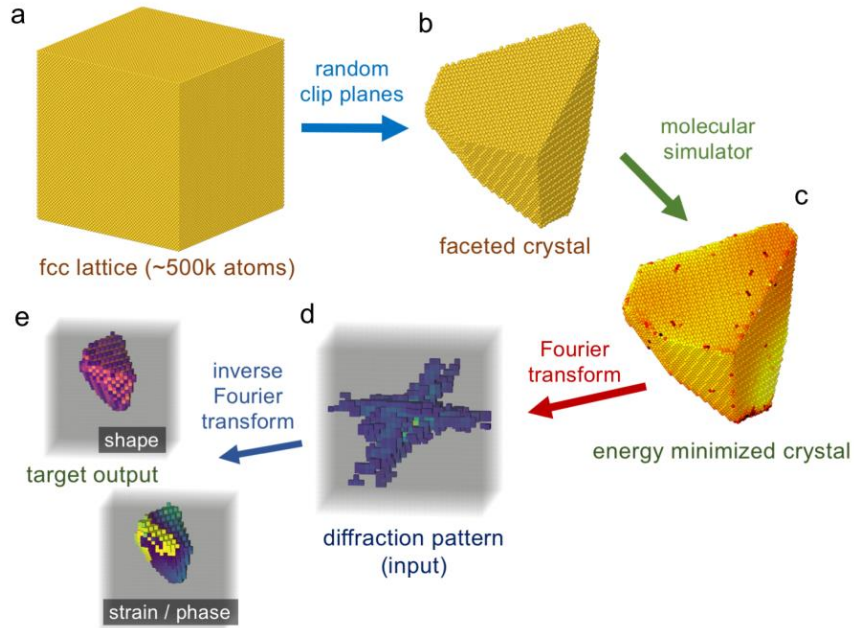

*Fig. 2: Preparation of physics informed training set for 3D-CDI-NN model using atomistic modeling. (a) A block of about half a million gold atoms arranged in fcc structure. (b) Atoms are removed from the fcc block using clip planes at high symmetry orientations that are randomly selected, which leads to gold crystal of random polyhedral. (c) A random combination of compression, tension and shear stresses is applied on the gold crystal followed by energy minimization using a molecular simulator to create a realistic strain field in the material. (d) 3D coherent diffraction pattern, i.e., input of the 3D-CDI-NN model, is prepared from downsampling of the electron density of the atomistic object followed by Fourier transform. (e) The corresponding real-space shape and phase of the object, i.e., target output of the 3D-CDI-NN model, is obtained via inverse Fourier transform of the 3D diffraction image.*

**Neural network architecture.** 3D-CDI-NN is a feed-forward neural network (Fig. 3) and consists of a convolutional autoencoder and two identically structured deconvolutional decoders. The encoder takes a 32×32×32 input image of 3D diffraction pattern magnitude and encodes it via a series of rectified linear unit (ReLu) convolutional layers and max pooling layers into a latent space that represents the underlying features. The same encoded data is passed through a series of ReLu convolutional layers and upsampling layers in two separate decoders to obtain 32×32×32 output images that map the encoded diffraction pattern to the corresponding shape and phase of the real-space image. A 3×3×3 kernel size is used as the convolution, max pooling, and upsampling windows. The network is trained in a supervised manner, where the output images for the training diffraction data are known a priori. In addition, the physics of the forward model is enforced via a custom objective function that minimizes the mean absolute error between the magnitude of input diffraction pattern and that obtained from Fourier transform of the recombined predicted shape and phase images. Dropout layers are added to the input layer and convolutional layers to help train a robust network. Dropout, which is the practice of randomly suppressing the output of various neurons during training helps to train more robust neural networks by forcing the network to learn multiple representations of the same input data.[27] The convolutional and max pooling operations (max pooling is a binning/downsampling operation using the maximum value over a prespecified pixel neighborhood) serve to transform the data (in this case the diffraction magnitudes) into feature space, while the deconvolutional and upsampling operations serve to transform back from feature space into image space.

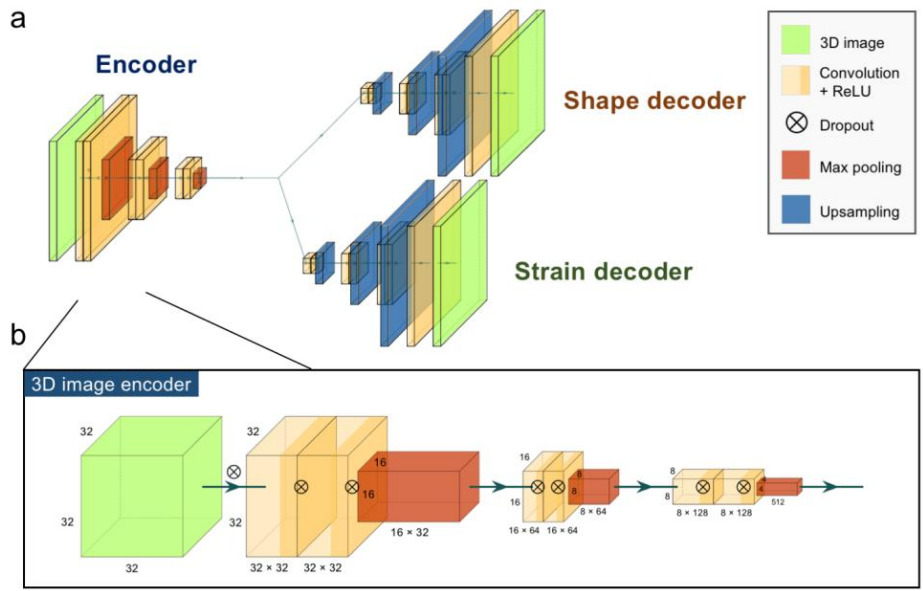

**Fig. 3: Schematic of the neural network structure of 3D-CDI-NN model.** (a) The model is a feed-forward network consisting of a convolutional autoencoder and two identically structured deconvolutional decoders. (b) The encoder takes a 32×32×32 input diffraction pattern and puts it into a feature space via a series of convolutional and max pooling layers (*i.e.*, input 32×32×32 →

16×16×64 → 8×8×128 → 4×4×256 latent space). Note that layers in the schematic are not drawn to scale. Identically structured decoders both take the same set of encoded features but respectively predict the shape and phase of the real space image. The decoding process is the reverse of the encoding process (*i.e.*, latent space 4×4×256 → 8×8×128 → 16×16×64 → 32×32×32 output) and it is achieved via a series of convolutional and upsampling layers. The physics of Fourier transform is enforced via a custom objective function that minimizes the mean absolute error between the magnitude of input diffraction pattern and that from the Fourier transform of the recombined predicted shape and phase images.

**3D-CDI-NN performance on simulated data.** Fig. 4 shows examples of the performance of 3D-CDI-NN on simulated data outside of the training dataset. From 32×32×32 input simulated diffraction patterns, 3D-CDI-NN predicts the corresponding real-space images in the same number of volume elements (voxels), i.e. for a 32×32×32 input diffraction pattern, 3D-CDI-NN makes 65536 predictions that correspond to the amplitude and phase in each voxel in the sample space. As seen in Fig.4, 3D-CDI-NN does a remarkable job of predicting sample structure and strain from input diffraction data alone. Although the predicted real-space images are pixelated due to the limited number of voxels, 3D-CDI-NN nevertheless predicts the facets and edges of the objects, without the need of any thresholding. Due to symmetry of the diffraction pattern, 3D-CDI-NN occasionally predicts the twin image of the target (*e.g.*, crystal 1 in Fig. 4) which is inverted in space and the complex conjugate in the phase of the target image. 3D-CDI-NN tends to overpredict phases of the real-space image when the object is weakly strained, which we partly resolved by adding examples of crystals with no strains to the training data (see Methods).

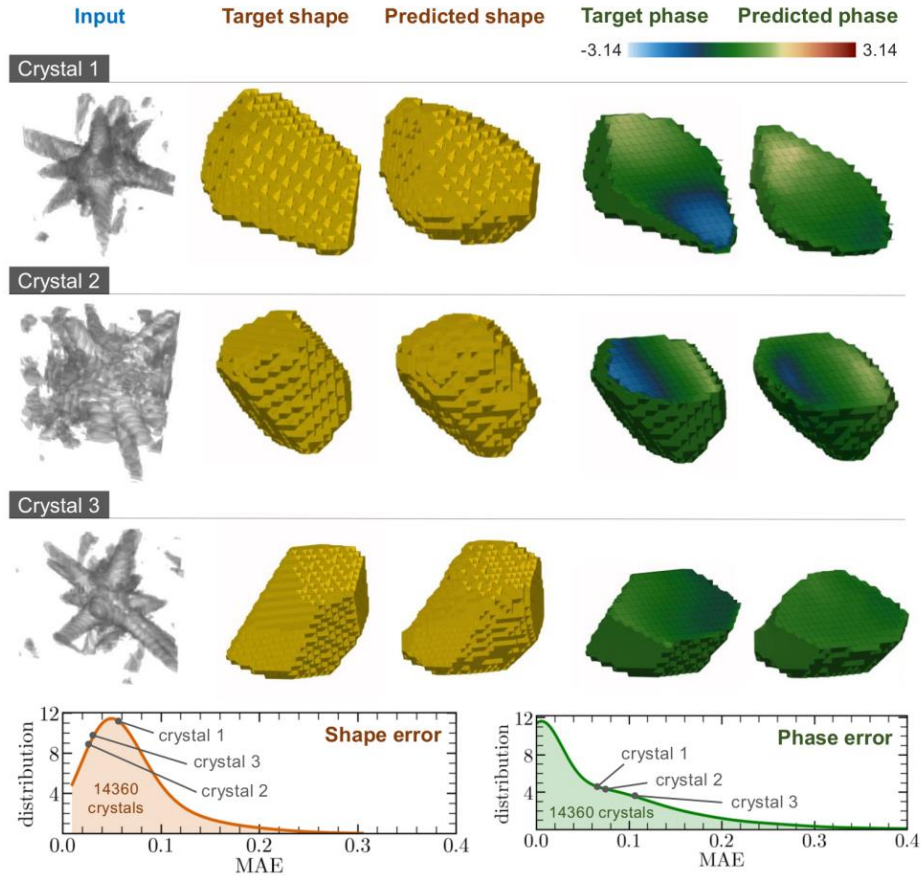

**Fig. 4: Performance of 3D-CDI-NN on simulated test data.** Three representative crystals randomly drawn from the test data set. For each crystal, we show the 3D input image, target, and predicted 3D images of the objects shape and strain fields. The phase images are clipped to show the internal strain fields. The plots show the normalized distribution of error in the predictions across the entire test dataset of 14360 simulated crystals (*i.e.*, crystals that are not used for training).

**Experimental BCDI measurement.** A sample containing gold nanoparticles on a Si substrate was prepared by dewetting a thin film of gold at 900 C. A nanoparticle was chosen at random and illuminated by a coherent beam focused to ~500 nm X 500 nm, which was large enough to fully illuminate the nanoparticle. We measured the resulting 3D coherent X-ray diffraction pattern about the crystals (111) Bragg peak. To obtain this 3D data set, we rotated the sample through 0.6 degrees in steps of 0.005 degrees, which resulted in 120 2D slices through the diffraction pattern. These slices were stacked in the 3$^{rd}$ dimension to give a data set of 151×133×103 reciprocal space voxels. The 55 micrometer pixels on the detector were sufficient to oversample the measured diffraction data by a factor of 2 or more at the detector distance of 0.9 m.

**Model validation on experimental data.** To evaluate the performance of the trained 3D-CDI-NN model on real data, we prepare input data by down-sampling 3D coherent diffraction pattern of the gold crystal obtained from the X-ray diffraction experiment. The down-sampling to 32×32×32 data is done via cropping and block-wise discrete cosine transform (*i.e.*, dct → cropping → inverse dct on blocks).[28] The target for comparison is prepared via tradition reconstruction of the original 3D coherent diffraction pattern followed by scaling, normalization, and binning to 32×32×32 images. Fig. 5 shows the performance and computational efficiency of the methods. 3D-CDI-NN model accurately predicts the shape and facets of the target crystal on a sub-second time scale (~145 milliseconds/prediction) but underestimates the crystal size and its local strain.

**DL Prediction refinement.** To improve the quality of reconstruction obtained through our approach, we refine 3D-CDI-NN's structure and strain prediction through an iterative gradient-based minimization procedure. We implement this refinement step within the same software package as the 3D-CDI-NN model (in our case Google's Tensorflow) by using the reverse-mode automatic differentiation (AD) technique; the use of the AD technique provides us with two pertinent advantages.[25] First, we only need to specify the physical forward model that describes the BCDI experiment, without having to actually derive the gradient expressions for both the physics-aware NN training and the final refinement steps. Second, we can directly use the sophisticated minimization algorithms present within the Tensorflow package instead of custom implementations. As seen in Fig. 5c, refinement of the 3D-CDI-NN prediction using AD recovers the crystal size and the inhomogeneous distribution of strain within the crystal. Benchmarking on the same CPU processor core shows that the combination of 3D-CDI-NN and AD is still ~4 times faster than traditional iterative phase retrieval method.

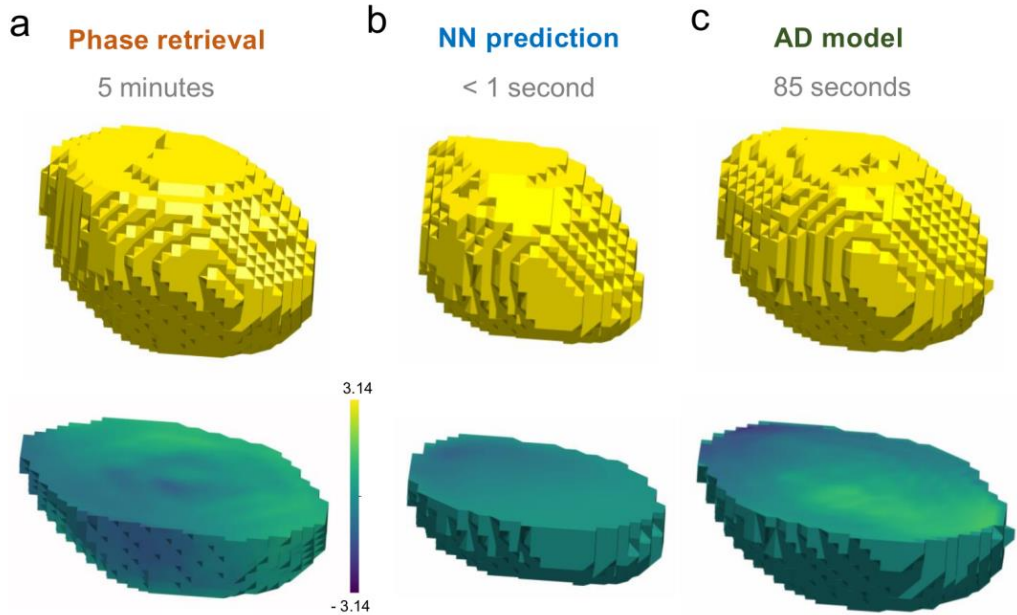

**Fig. 5: Validation of 3D-CDI-NN model on real image data from CDI experiments.** Comparison between reconstructions from traditional phase retrieval, 3D-CDI-NN prediction, and AD refined 3D-CDI-NN prediction.

## Discussion

In conclusion, we have demonstrated for the first time the use of machine learning to invert 3D coherent imaging data rapidly and accurately. Once trained, 3D-CDI-NN is hundreds of times faster than traditional iterative phase retrieval methods. While 3D-CDI-NN demonstrates excellent performance on simulated test data, there is much scope to improve its performance on experimental test data. We expect this gap in performance can be addressed in several ways, including through transfer learning and neural architecture search. Transfer learning is a powerful means of training large neural networks in the absence of sufficient amounts of training data. The neural network is first pre-trained using a large data set on a similar problem before being refined using the smaller data set corresponding to the target problem. We can apply the same method to 3D-CDI-NN by pre-training on simulated data, before refining its training on experimentally phased data sets. We expect this new network to significantly perform better on fresh experimental data. Another important means of improving network accuracy which we have not explored is by optimizing the architecture of the network (which was hand-engineered and kept fixed for this study) (Fig. 3). Automated approaches to neural network design are now widely used and can generate network architectures that surpass the best designed human ones.[29,30]

We anticipate that modern data analytical approaches to coherent diffraction inversion will be critical to CDI at the coming fourth generation synchrotron sources, such as the recently commissioned Extremely Brilliant Source (EBS) at the European Synchrotron Radiation Facility and the coming Advanced Photon Source Upgrade. At these sources, the coherent flux of the beam is expected to increase by a factor of 50 – 200 times over current sources. This vast increase in flux can be used to measure both higher resolution data (corresponding to much larger data sets), or measure at current resolutions but at significantly higher rates, just tens of seconds per measurement.

The current phase retrieval methods will not keep up, in either larger data sets or faster data rates due to limitations of modern GPU devices in both compute cores and memory. In fact, our 3D-CDI-NN approach has shown to produce high fidelity images from very limited data. This image is then refined through a gradient-based minimization procedure. In the case of extremely large data, which precludes full iterative phase retrieval on current GPU devices due to limited onboard memory, we envision 3D-CDI-NN being used to initialize an AD solution that is based on the entire data volume. Additionally, since CDI-NN is not doing phase retrieval, the typical oversampling of reciprocal space is not required for inversion to real-space images. The neural network can be used on only a limited volume of data, perhaps very close to a Bragg peak of the lattice. One will then optimize the measurement-inversion process to have the neural network working on the subset of data before the total volume of data is even finished acquiring. AD has been shown to scale effectively to large compute resources, contrary to current phase retrieval algorithms, and can even operate on the partial data set as it is being acquired.

## Methods

**Atomistic model of gold crystals.** Each gold crystal is cut from a ~ 20 nm × 20 nm × 20 nm face center cubic (fcc) lattice of ~ 500k atoms, where the direction (normal vector) of each clip plane is uniformly sampled using the hypersphere point picking method. A random number of selected high symmetry orientation clip planes (between $4 - 20$) positioned at random distances from the crystal geometric center is used to obtain faceted gold crystals of diverse shapes and sizes. The crystal is minimized in LAMMPS using the embedded-atom method (EAM) interatomic potential to obtain the initial gold crystal structure. The final gold crystal structure is obtained by applying a combination of compression, tension, and shear stresses (up to 1% strain) to the initial structure followed by another minimization. The stresses are applied to the structure via deformation of the simulation box with atom coordinates remapped accordingly. Both the initial and final structure of gold crystal are scaled by the inverse lattice constant of gold (1/4.078 Å) such that the lattice constant is normalized to 1. The final structure is used to compute the crystal shape whereas the difference between the initial and final structures is used to compute the crystal phases (see Training data below). To avoid potential artifact from boundary related problems, a ~ 5 lattice unit padding (*i.e.*, ~ 20 Å before lattice normalization) is added to each side of the normalized (lattice constant = 1) simulation box.

**Training data.** The training dataset is a combination of two datasets. The first dataset consists of 107,180 diffraction patterns generated from atomistic models of gold crystals, where 100,000 of them are used for training and 7180 are set aside for testing. The second dataset consists of the same 100,000 and 7180 gold crystals in the training and testing sets but with the material strain field removed (*i.e.*, the testing set remains entirely independent from the training set). The second dataset serves as control that helps alleviate the tendency of 3D-CDI-NN in overpredicting strains. Target output images of the crystal shape is obtained from the number density of atoms calculated using a bin size of ~ 2 lattice units (~ 8 Å before normalization) and normalized by the maximum whereas target output images of the crystal phases is obtained from binning of local phases that is computed from the atom displacement field of the final and initial crystal structure projected along [111] and scaled by $2\pi$. The binning process convert atomistic model into 32×32×32 images (*i.e.*, length of each voxel corresponds to ~ 7.5 Å before lattice constant normalization). For each crystal, the shape (magnitude) and phase images are combined to form a 3D array of complex numbers which is then used to obtain the corresponding diffraction pattern via Fourier Transform. The magnitude of the 3D diffraction pattern is used as the input for the 3D-CDI-NN training.

**3D-CDI-NN training.** Training was performed in parallel on 4 NVIDIA V100 GPUs using the Keras package running the Tensorflow backend.[31,32] We trained the networks for 50 epochs each using a batch size of 256. The training for each network took less than half an hour when trained in parallel across the 4 GPUs. At each step, we used adaptive moment estimation (ADAM)[33] to update the weights while minimizing the per-pixel mean absolute error. A 10% dropout rate is applied to all dropout layers. We computed the performance of the network at the end of each training epoch using the validation set. Since the network architecture of the 3D-CDI-NN model consists of a common encoder shared by two separate decoders, we adapted a systematic approach in training the model weights. We first trained the encoder and shape decoder. We subsequently fixed their weights while performing the initial training of the phase decoder. This was followed by a further training step which involved unfixing the encoder weights and additional training of the phase decoder. The final step was the simultaneous refinement in the weights of all branches of the network. We found that this sequential training approach was necessary to stabilize a network involving multiple branches which tends be unstable (fluctuate wildly) in the beginning due to random initialization of weights and the inability of a single weighted sum objection to handle the case where improvements in one branch is canceled by other branches.

**Iterative phase retrieval.** To perform phase retrieval, we used standard iterative phase retrieval algorithms that switched between error reduction (ER) and hybrid input-output (HIO).[34] 620 iterations were performed using a shrink-wrapped support in real space.[17] The final 20 iterations were averaged over to obtain the final result. The only difference in the phased data was that oversampling was required so the DCT down sampling was not performed.

## Data Availability

The trained network, test data and accompanying Jupyter notebooks of Python code are available upon reasonable request to the corresponding author.

# Acknowledgements

This work was supported by Argonne LDRD 2018-019-N0 (A.I C.D.I: Atomistically Informed Coherent Diffraction Imaging). An award of computer time was provided by the Innovative and Novel Computational Impact on Theory and Experiment (INCITE) program. This work was performed, in part, at the Center for Nanoscale Materials, a U.S. Department of Energy Office of Science User Facility, and supported by the U.S. Department of Energy, Office of Science, under Contract No. DE-AC02-06CH11357. This work also used computing resources provided and operated by the Joint Laboratory for System Evaluation (JLSE) at Argonne National Laboratory. Use of the Advanced Photon Source and Argonne Leadership Computing Facility, both Office of Science user facilities, was supported by the U.S. Department of Energy, Office of Science, Office of Basic Energy Sciences, under Contract No. DE-AC02-06CH11357.

# Author contributions

M.J.C., H.C, Y.N, S.S and R.H designed the research. H.C built, trained, and tested the deep learning networks. S.K., S.H. and Y.N. contributed to the refinement of the network predictions through AD. All authors contributed to the analysis, discussion and writing of the manuscript.

# Competing interests

The authors declare no competing interests.